\documentclass[usenatbib, twocolumn]{mnras}
\usepackage{mathtext,amssymb,amsmath}
\usepackage{epsfig}
\usepackage{graphicx}
\usepackage[export]{adjustbox}
\usepackage{url}
\usepackage{upgreek}
\usepackage{comment}
\usepackage{times}
\usepackage[T1]{fontenc} 
\usepackage{aecompl}

\renewcommand{\vector}[1]{\ensuremath{\mathbf{#1}}}

\renewcommand{\div}{\ensuremath{..}}

\newcommand{\bea}{\begin{eqnarray}}
\newcommand{\eea}{\end{eqnarray}}

\newcommand{\ergl}{\ensuremath{\,\rm erg\, s^{-1}}}

\newcommand{\pardir}[2]{\ensuremath{\frac{\partial #2}{\partial #1} }}

\newcommand{\bqcam}{V\,0332+53\,}
\newcommand{\gro}{GRO\,1744-28}

\renewcommand{\div}{\ensuremath{-}}

\title[Inertial oscillation modes in a magnetosphere]{Inertial oscillation modes of an inclined dipolar magnetosphere as a source of band-limited noise in X-ray pulsars}
\author[P. Abolmasov and A. Biryukov]{Pavel Abolmasov$^{1,2}$\thanks{pavel.abolmasov@gmail.com} and Anton Biryukov$^{2,3}$\\
$^{1}$Department of Physics and Astronomy, FI-20014 University of Turku, Finland\\
$^{2}$Sternberg Astronomical Institute, Moscow State University, Universitetsky pr., 13, Moscow 119234 \\
$^3$ Kazan Federal University, Kremlyovskaya str. 18, Kazan, 420008, Russia
}

\begin{document}

\date{Accepted ---. Received ---; in
  original form --- }

\label{firstpage}
\pagerange{\pageref{firstpage}--\pageref{lastpage}} \pubyear{2016}
\maketitle

\begin{abstract}
Magnetic fields of strongly magnetized stars can trap conducting matter due to frozen-in condition. In the force-free regime, the motion of the matter along the field lines
may be considered in the ``bead on a wire'' approximation. Such a motion, if gravity and centrifugal forces are taken into account, has equilibrium points, some of which are stable. In most cases, stability is possible in about several per cent of the possible locations. Corresponding oscillation frequencies span the range from zero to $\sqrt{3}$ of the spin frequency. We suggest that this variability mode may be excited in some X-ray pulsars during the outbursts and create the peaked broad-band noise component near the break frequency in the power density spectrum, as well as produce some of the quasi-periodic oscillation features in this frequency range.  Existence of this variability does not require any changes in mass accretion rate and involves only a small amount of matter infiltrating from the disc and magnetic flow due to interchange instabilities. 
\end{abstract}


\section{Introduction}

A classical picture of accretion upon a magnetized neutron star (NS) involves an accretion disc, truncated by the magnetic forces at its inner boundary, and a flow inside the magnetosphere, channelled along the magnetic field lines towards the surface of the star.
Most of the energy release occurs near the surface of the star in a narrow region (the hotspot).
The local Eddington limit in the hotspot may be easily overcome due to its relatively small size, forming an accretion column, supported by radiation pressure forces \citep{BS76}.

Such a model has its imprints upon the variability of X-ray binaries.
In particular, accretion discs in general are predicted to be sources of a broad-band noise \citep{lyubarski} truncated at either the maximal viscous (as in the original studies) or at the highest dynamical (or, more precisely, the vertical breathing mode, as recent numerical simulations suggest, see for example \citealt{misra19}) frequency.
This leads to a break in the power density spectrum (PDS).
Indeed, in many X-ray pulsars, there is a break at a frequency close to the spin frequency of the NS, slightly variable with the varying mass accretion rate \citep{revnivtsev09}. 
However, the motion of the break is not the only effect: during the outbursts, many X-ray pulsars develop an excess broad-band noise component around the break frequency. 

Does the magnetosphere itself produce any types of variability?
This is a question largely understudied, both in the observational and in theoretical works.
The only exception is the high-frequency (tens of Hz) variability of the ``bursting pulsar'' \gro\ \citep{klein96} interpreted as a consequence of photon-bubble instability in the accretion column.

Apart from the magnetic field whose pressure dominates over gas pressure within the magnetosphere, matter within the magnetosphere exists in an effective gravity field weakened by radiation pressure. This allows parts of the material in the magnetosphere to reach local equilibrium between rotation and effective gravity.
As we show in this paper, under certain circumstances, this equilibrium may be stable.  Stable equilibrium points allow for periodic small-amplitude oscillations. 
Here, we consider the properties of these oscillations. 

Of course, to reach the equilibrium points, matter needs to get onto the closed lines of the magnetosphere. 
And this is effectively a part of the larger case of magnetospheric accretion, because accreting matter also needs to get inside the magnetosphere. 
Accretion upon magnetized NSs is a complicated issue in which not all the details are yet clearly understood. Evidently, some of the magnetic field lines of an accreting magnetized NS are open, and some are closed, and the interaction with the disc tends to open field lines by bending them \citep{AK90, lovelace95, uzdensky02}. 
At the same time, because of the very strong dependence of magnetic field pressure on radius and of the essentially infinite conductivity of the plasma, the interaction between the disc and the NS magnetic field is possible in a narrow range of radii \citep{KR07,CAP17}. 
The magnetic field lines of the star expected to pierce the disc at larger distances are inevitably open (see also \citealt{parfrey16}), and the closed lines that do not reach the interaction zone are essentially in an unperturbed regime. 
Neglecting higher multipoles of the NS field, we can safely treat the field as dipolar. 
We know that even fully ionized matter is capable of settling onto the closed field lines via interchange instabilities \citep{AL76} and by reconnections between the magnetic fields frozen into the matter of the disc and the dipolar field of the star.
 This mechanism is often involved to explain the accretion flows inside the magnetosphere. 
Interchange instabilities effectively provide diffusion of the matter from the disc and populating the closed field lines.
However, the amount of mass loading onto a particular field line varies strongly, meaning that even at a very high mass accretion rate some field lines have very little mass on them. 

All the calculations in this paper are non-relativistic, that is a reasonable approximation for a large, slowly rotating magnetosphere of an X-ray pulsar. 
This assumption is true for the radii much larger than the size of the NS but much smaller than the light cylinder radius. 
Periods of most X-ray pulsars range from seconds to tens and hundreds of seconds (see for example \citealt{raguzova}), meaning that the light cylinder is at least a couple of orders larger than the corotation radius (and thus the expected radius of the magnetosphere). 
The large value of the light cylinder radius also means that the deformation of the magnetic field due to rotation is negligibly small.
At the same time, the magnetosphere, as it was mentioned, is much larger that the size of the NS, that allows to neglect all the higher multipoles of the magnetic field of the star, as their relative contribution to the field rapidly decreases with radius. 

Another, and perhaps the most restrictive, assumption we will make here is the force-free approximation: we will assume that in the considered portions of the magnetosphere magnetic stresses dominate over pressure and inertia of the oscillating matter. 
All these conditions ensure that, in the reference frame co-rotating with the NS, magnetic field is a pure untwisted dipole, and the plasma moves along the field lines only. 
As we will show in Sect.~\ref{sec:calc:just}, if the pressure gradients along the field lines are ignored, this leads to the ``bead on a wire'' approximation \citep{HR71,BP81}. 

In Section~\ref{sec:calc}, we calculate the positions of the equilibrium points, find the conditions for stability, and estimate the oscillation frequencies in the stable regions.
In Section~\ref{sec:disc}, we discuss the limitations of the approximations used, the possibilities for observing these oscillation modes, and apply the results to the variability modes observed in X-ray pulsars.
We conclude in Section~\ref{sec:conc}.

\section{Motion inside a dipolar magnetosphere}\label{sec:calc}

\subsection{Justification for the kinematic approach}\label{sec:calc:just}

In the most general, covariant, formalism, the equation of motion has the form
\begin{equation}\label{E:arel}
    \rho a_i = \frac{1}{4\uppi} \left( \left[\nabla \times \vector{B} \right]\times \vector{B}\right)_i - P_{; i} + f_i,
\end{equation}
where $a_i = u^k u_{i;k}$ is 4-acceleration (also including gravity and inertial forces), $\rho$ is co-moving rest-mass density (it is safe to ignore the inertia of heat as long as the temperature of the flow is non-relativistic), and the right-hand side contains the contribution from magnetic fields ($B^i$ is a 3-vector, and the contribution from electrostatic field is negligible in the co-rotating frame), radiation pressure 4-force $f_i$, and pressure gradient. Semicolon is used for covariant derivative (see for example \citealt{MTW} Chapter 8). 

If the magnetic field dominates over pressure and inertia, transverse components of the magnetic force restrict the motions perpendicular to the field. 
At the same time, projecting the spatial part of (\ref{E:arel}) onto the magnetic field will imply
\begin{equation}
    \rho a_\parallel = f_\parallel - \nabla_\parallel P,
\end{equation}
as the projection of the magnetic force is exactly zero. 
For the pressure contribution to become important, pressure gradient should be comparable to gravity and centrifugal forces. 
This will require the gas to have temperature of the order virial, that is extremely unlikely as local synchrotron and non-local Compton cooling times are in general shorter than the dynamical time scales of the problem. 
Relevant estimates were made, for example, in \citet{shakura13}: for an X-ray pulsar flare with a luminosity $\gtrsim 10^{38}\ergl$, equation (34) of that paper predicts the Compton cooling time scale shorter than tenth of a second, that is evidently smaller than the dynamical time. 
At the same time, unlike accretion flows, there is no energy source heating the plasma up to virial temperatures.
This means that gas pressure may be ignored not only in comparison with the magnetic field stresses but also with respect to the inertia. 

Finally, the equation of motion is affected by radiation pressure force only, $a_\parallel = f_\parallel$. 
In the non-relativistic regime, this motion may be described as a motion in an effective potential including radiation, gravity, and centrifugal forces. 

\subsection{Kinematic model for a particle}

We will restrict ourselves to a simple ``bead on a wire'' approximation, in which the magnetic field plays a role of a kinematic constraint: a particle can move only along the field line. 
The field, in a co-rotating frame, has an exact unperturbed dipole shape. 
The reference frame used is aligned with the field and rotates with the star at a fixed rate set by the spin frequency $\Omega$.
Motion of the particle is essentially a motion in an effective potential
\begin{equation}\label{E:Phi}
    \Phi = - \left( 1-\Gamma\right) \frac{GM}{R} - \frac{1}{2} \Omega^2 R^2\sin^2\alpha,
\end{equation}
where $R$ is the spherical radial coordinate, $M$ is the mass of the star, and $\alpha$ is the angle between the radius vector and the rotation axis, and we also included the effect of radiation pressure through Eddington multiplier $\Gamma$. 
Such a correction may arise if a radiation source such as accretion column is present near the surface of the star. 
For the general case of an inclined dipole, angle $\alpha$ does not coincide with the polar angle $\theta$. If $\chi$ is the magnetic angle (the angle between the rotation and magnetic axes), they are related through the cosine theorem
\begin{equation}\label{E:alpha}
\cos\alpha = \cos\theta \cos\chi + \sin \theta \sin \chi \cos\varphi,
\end{equation}
where $\varphi$ is azimuthal angle of the spherical coordinate system. 

Position of an equilibrium point is given by condition $\partial \Phi / \partial L =0$, where $L$ is the coordinate along the field line.
For a given  field line, $L$ is a monotonic function of $R$ between 0 and the outermost radius.
For convenience, let us consider normalized quantities $l = L/R_0$, $x=R/R_0$, $\omega = \Omega R_0^{3/2} / \sqrt{(1-\Gamma)GM}$, and $\phi = \frac{\Phi}{1-\Gamma} \frac{R_0}{GM}$. 
Normalization radius $R_0$ may be set arbitrarily.

Potential in dimensionless units is now
\begin{equation}
    \phi = -\frac{1}{x} - \frac{1}{2}\omega^2 x^2\sin^2\alpha. 
\end{equation}
The equilibrium points are the zeros of the first derivative $d\phi / dl$, and their stability depends on the sign of the second derivative $d^2\phi / dl^2$. 

\subsection{Equilibrium points}

\begin{figure*}
\adjincludegraphics[width=\textwidth,trim={0 0 0 0},clip]{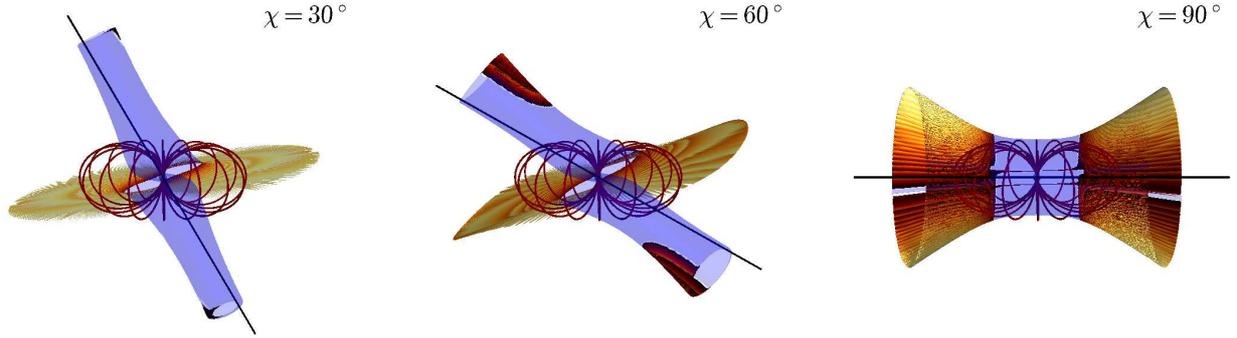}
 \caption{ Three-dimensional rendering of the surface of equilibrium points for $\chi = 30$, $60$, and $90\degr$, and $\Gamma = 0.9$. 
 Unstable part of the surface is shown as transparent blue, stable part as opaque yellow-red with a striped pattern. 
 Dipole magnetic field lines with the outermost radius corresponding to the classical corotation radius are shown with red lines. The black straight line shows the rotation axis.
 Position of the camera used for rendering is always at the azimuth of 90 degrees (while the rotation axis is at the azimuth of 0) and elevation of 10 degrees above the magnetic equator. 
 }\label{fig:xsurface}       
\end{figure*}

Directional derivative along the field
\begin{equation}
  \frac{d\phi}{dl} = \frac{dx}{dl} \pardir{x}{\phi} + \frac{d\theta}{dl} \pardir{\theta}{\phi},
\end{equation}
where the derivatives $d(x, \theta)/dl$ are calculated along a single field line,
\begin{equation}
    \frac{dx}{dl} = \frac{2\cos\theta}{\sqrt{1+3\cos^2\theta}},
\end{equation}
\begin{equation}
    \frac{d\theta}{dl} =  \frac{\sin\theta}{x\sqrt{1+3\cos^2\theta}}.
\end{equation}
For the first derivative, we get
\begin{equation}
\begin{array}{l}
  \displaystyle  \frac{d\phi}{dl} = \frac{1}{\sqrt{3\cos^2\theta+1}} \left[ \frac{2\cos\theta}{x^2} - \omega^2 x \left( \left( 3\sin^2\alpha - \sin^2\chi \right) \cos\theta \right.\right.\\
    \displaystyle  \left.\left. \qquad{}+ \sin\chi \cos\chi \cos \varphi \sin\theta\right)\right].\\
    \end{array}
\end{equation}
Equilibrium is reached, in a direction set by $\theta$ and $\varphi$, at a dimensionless radius
\begin{equation}\label{E:xeq}
  x = 2^{1/3}\omega^{-2/3} \left( 3\sin^2\alpha - \sin^2\chi + \sin\chi \cos\chi \cos\varphi \tan\theta \right)^{-1/3}.
\end{equation}
In particular, for an aligned dipole, the equilibrium (propeller) surface has the shape $x\propto \sin^{-2/3}\theta$ (see for example chapter III of \citealt{lipunov_book}). 
For arbitrary inclination, there are directions where no positive solutions for $x$ exist.
Sample shapes of the equilibrium surfaces for different magnetic angles are shown in Fig.~\ref{fig:xsurface}.

Note that for completely aligned case ($\chi = 0$), equilibrium radius in the plane of the NS equator ($\alpha = 90^{\circ}$) becomes $(2/3)^{1/3}\omega^{-2/3}$, which does
not coincide with the co-rotation radius $\omega^{-2/3}$. 
Apparent contradiction is related to the fact that the motions we consider are not free Keplerian motions in the gravity field but instead forced by the kinematic constraints. 
Corotation radius $R = \left(GM/\Omega^2\right)^{1/3}$ still remains a characteristic radius scale. However, in the presence of a radiation field, this quantity is altered by a factor $(1-\Gamma)^{1/3}$, that allows to shift the equilibrium points inside the magnetosphere.

\subsection{Stability and oscillation frequencies}

Second-order derivative of the effective potential requires much more computations, but all of them are straightforward.
In a relatively compact form, it can be written as
\begin{equation}\label{E:ddf}
\begin{array}{l}
\displaystyle 
    \frac{d^2\phi}{dl^2} = \frac{1}{\left(1+3\cos^2\theta\right)^2}\left[ 
    -\frac{2}{\left| x\right|^3} \left(1+3\cos^2\theta + 12\cos^4\theta\right)  \right. \\
    \left. \qquad+ \omega^2 \left( \left(3\sin^2\alpha - \sin^2\chi\right) \left(1-3\cos^2\theta - 6 \cos^4\theta \right) \right.\right. \\
\displaystyle 
    \qquad{}\left.\left.  + 6 \sin^2\theta \cos^2\theta (1+3\cos^2\theta) \left(\sin^2\chi\cos^2\varphi - \cos^2\chi\right)\right.\right.\\ 
\displaystyle 
\left.\left. \qquad{}    - 12 \sin\chi \cos\chi \cos\varphi \sin\theta \cos\theta \left( 1+\cos^2\theta -3\cos^4\theta \right)\right)
    \right]. \\
    \end{array}
\end{equation}
If the absolute value of $x$ is used in this expression, it also remains valid in the negative effective mass case  ($\Gamma>1$), see Section~\ref{sec:negm}. 
Near an equilibrium point, a test particle is pulled by a restoring force $F = -\varkappa^2 (l-l_{\rm eq})$, where $\varkappa^2 = d^2\phi/dl^2$, if positive, is the oscillation frequency in the vicinity of the point.
The result scales with $\omega^2$, as the equilibrium points we consider here always involve centrifugal forces.
Negative values of the second derivative mean unstable equilibrium points.
It is easy to check that for an aligned dipole all the equilibrium points are unstable. 

\begin{figure*}
\includegraphics[width=0.9\textwidth]{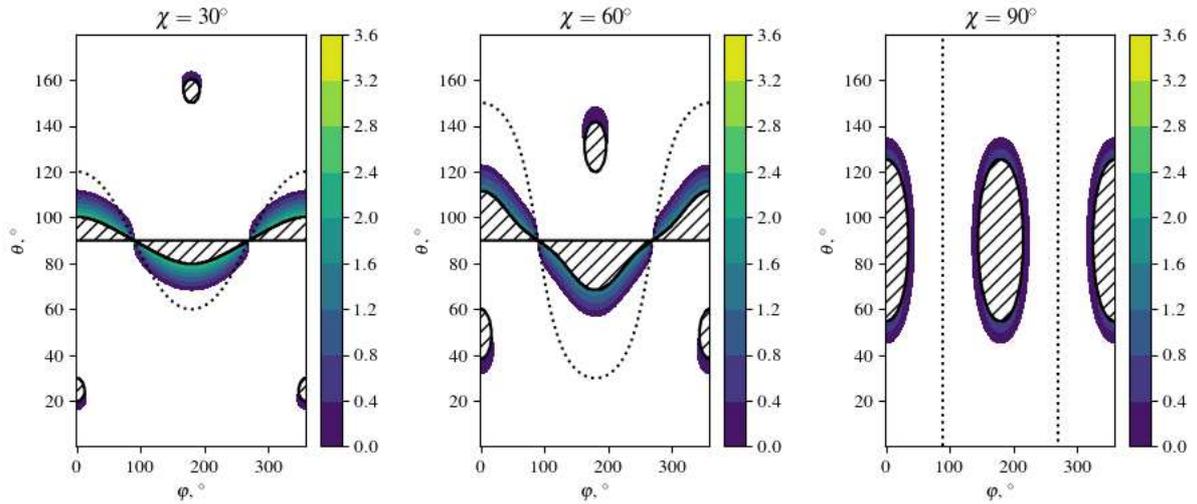}
 \caption{ Direction maps for rotating dipoles with $\chi = 30$, $60$, and $90\degr$. Shaded regions have equilibrium points with positive $\varkappa^2$ (colour-coded in $\omega^2$ units). Hatched regions have no equilibrium points at all. Dotted curve is the rotational equator ($\alpha = \pi/2$). 
 }\label{fig:oscontours}       
\end{figure*}

\begin{figure}
\includegraphics[width=0.9\columnwidth]{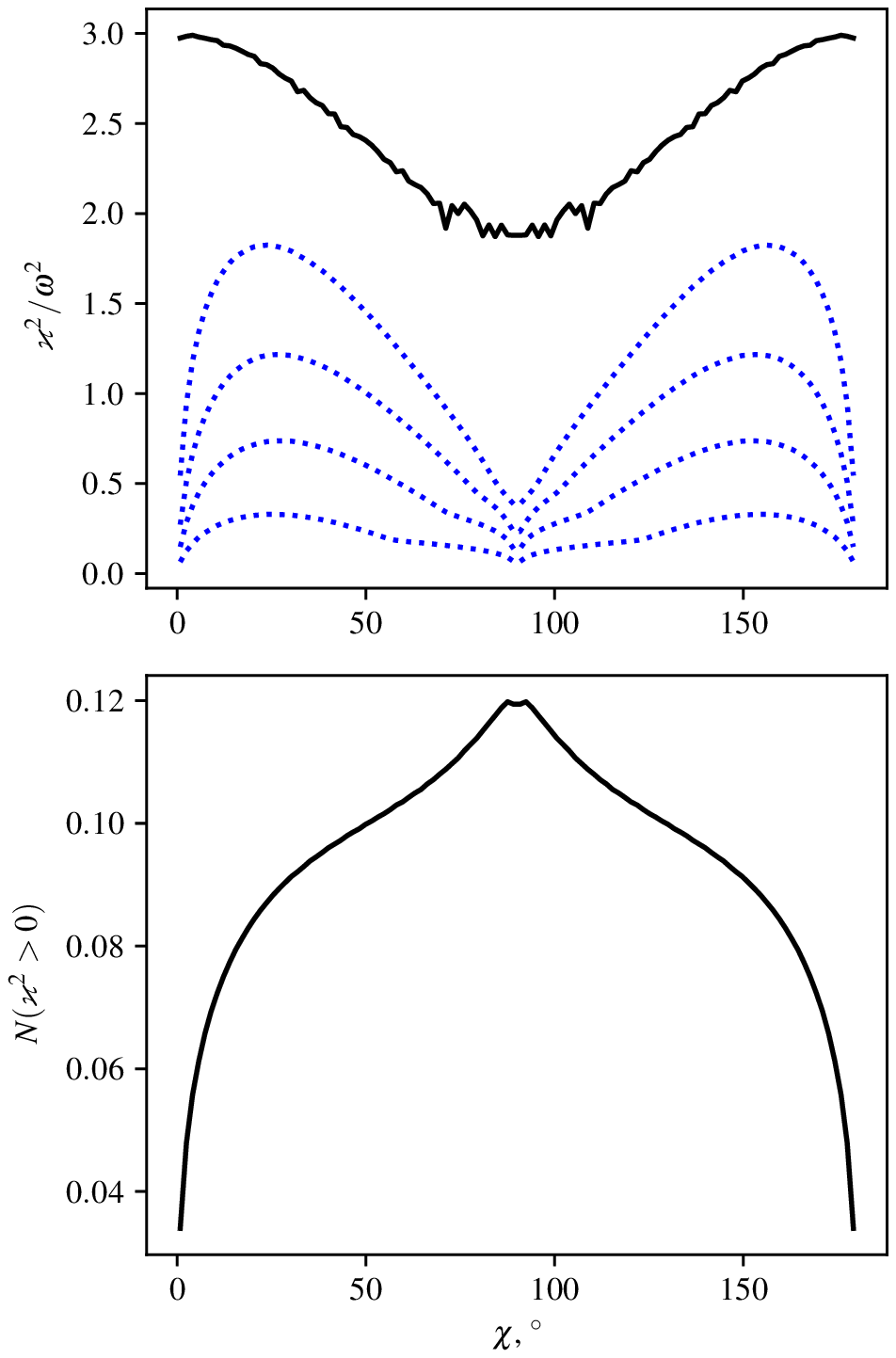}
 \caption{ Upper panel: maximal epicyclic frequency squared  as a function of magnetic angle. Dotted blue lines are frequency quantiles (0.2, 0.4, 0.6, and 0.8). Lower panel: fraction of the total solid angle where $\varkappa^2>0$.  The solid angle is calculated from the point of view of an observer located at the center of the NS. 
 }\label{fig:ostats}       
\end{figure}

In Fig.~\ref{fig:oscontours}, we show several maps of $\varkappa^2$ in the $\theta$-$\varphi$ space.
We also mark the points where there are no equilibrium solutions.
For an arbitrarily inclined dipole with $\sin\chi \neq 0$, there are always directions where $\varkappa^2$ is positive in the equilibrium point.
The fraction of such points is about several per cent (up to 12 per cent for an orthogonal rotator). For small magnetic angles, they mostly concentrate in a narrow belt situated between the magnetic equator 
and the rotational equator, and also in crescent-shaped regions close to the rotation axis on the side of the closest magnetic pole.  These secondary stable regions are usually located far away from the NS, but approach at higher $\chi$ (see Fig.~\ref{fig:xsurface}).
The maximal possible frequency is about $\sqrt{3}\omega$.
In Fig.~\ref{fig:ostats}, we show how the frequency distribution changes with  magnetic angle.

\subsection{Negative effective mass}\label{sec:negm}

Though one would not expect any astrophysical objects, especially accreting, to have negative masses, strong, compact radiation source situated near the surface of the star may mimic this exotic situation. 
Real super-Eddington sources, of course, should have finite size, but we so far limit ourselves to a simple, point-source, approximation. 
Even then, the contribution of the radiation source depends also on the optical depth $\tau$ of the test fluid particle: in the optically thin case, violation of the Eddington limit means effectively negative mass, while taking into account the optical depth effects alters the Eddington factor $\Gamma$ by about a factor of $\tau$, as the same radiation pressure force is now applied to a larger mass. 
In this case, the sign of equation~(\ref{E:xeq}) should change (let us consider the power of $1/3$ as a cubic root), but the expression for the oscillation frequency~(\ref{E:ddf}) remains valid. 

\begin{figure}
\includegraphics[width=0.9\columnwidth]{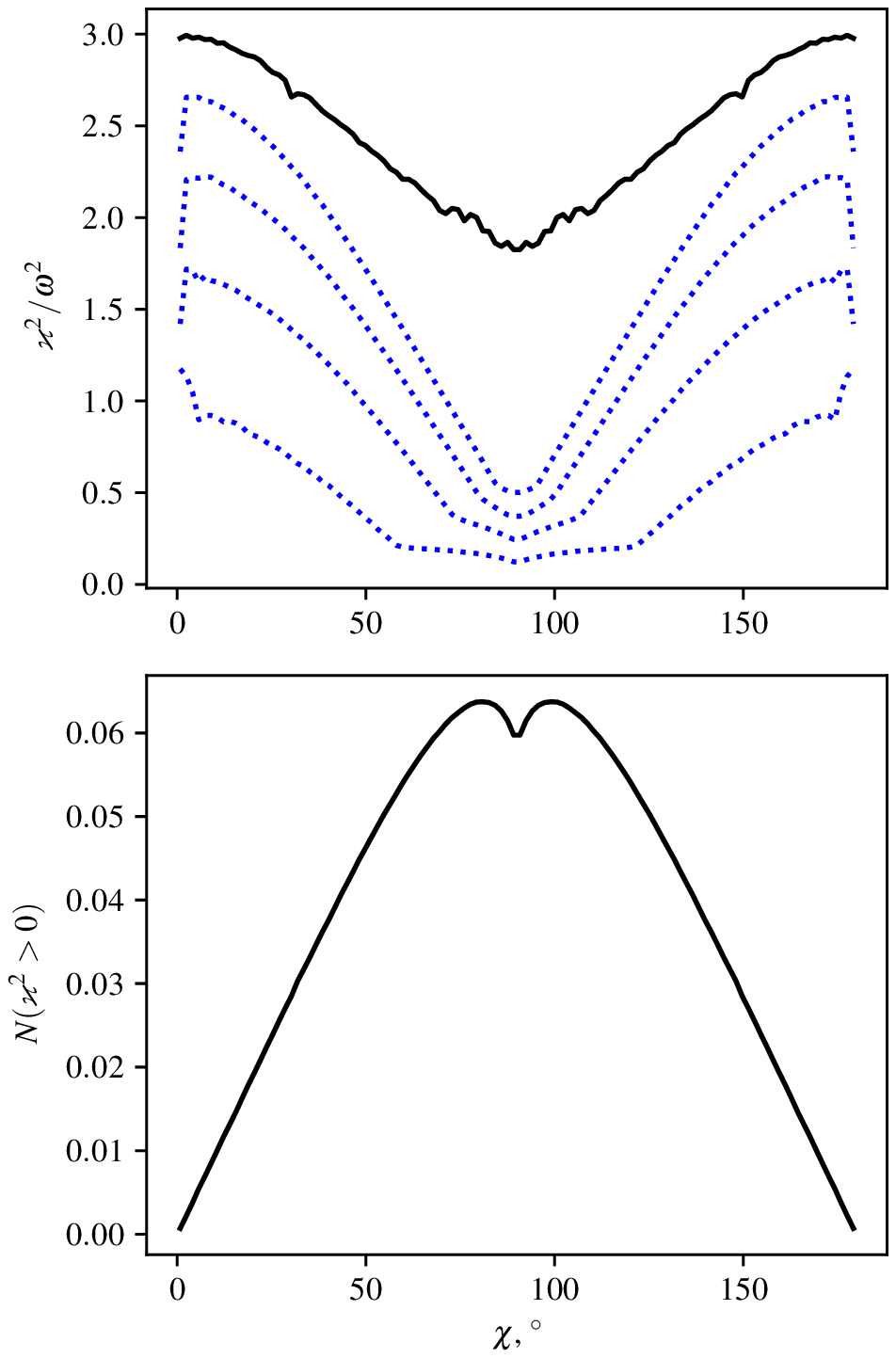}
 \caption{ The same as the previous figure, but for the negative effective mass case.  
 }\label{fig:ostats_negm} 
\end{figure}

Qualitatively, the spatial distribution of the stable zones remains similar, but now they are confined to the former regions of negative $x$.
The fraction of stable points is smaller, up to 6 per cent (see Fig.~\ref{fig:ostats_negm}). 
As in the positive-mass case, the characteristic frequencies form a broad distribution in the range $\left(0\div \sqrt{3}\right)\Omega$. 

\section{Discussion}\label{sec:disc}

\subsection{Properties of the oscillations}

When the inertial oscillations are possible, their frequencies span the range from 0 to a maximal value dependent on $\chi$ and changing between approximately $\sqrt{2}\Omega$ (for an exactly orthogonal dipole, the maximum frequency value is slightly smaller than $\sqrt{2}$ times spin frequency) and $\sqrt{3}\Omega$. 
Distribution over the frequencies is smooth, with a median value close to $\Omega$. 
Thus, it is impossible to explain the frequencies larger than a couple of spin frequencies with inertial modes in the approximation we use. 
Whenever centrifugal force is important, it sets the frequency scale. Hence, independently of the size of the magnetosphere and the contribution of radiation pressure to effective gravity, the characteristic frequencies of the inertial modes are close to the spin and create excess power around the spin frequency. 
Depending on the oscillation excitation conditions and the presence of matter in the regions of stable equilibrium, the actual variability spectrum may vary, but the distribution of frequencies shown in Fig.~\ref{fig:ostats} suggests roughly constant power up to about $\left( 0.5\div 1.0\right) \Omega$ with a gradual decline at higher frequencies. 
Such spectral component is harder than the flickering spectrum of the disc with the power roughly inversely proportional to frequency \citep{lyubarski}, and thus will stand out only near the break. 

This picture is similar to the effects observed in some X-ray pulsars in outbursts (see for instance fig.~2 of \citealt{revnivtsev09}, or \citealt{monkkonen19}), where the flicker noise becomes contaminated with a peaked (in the sense of $f\times PDS$) component centered at about the break frequency. 
Though the presence of a peak close to the break frequencies is questionable (the same excess in spectral power may be explained by a harder flicker noise slope), some X-ray pulsars also show QPOs at frequencies comparable to the break frequency \citep{angelini89, finger96, RN13}.
As those features are usually observed during outbursts, and the PDS in quiescence are much flatter, increase in luminosity is apparently an important factor. 
For a NS close to or exceeding its Eddington limit, the $1-\Gamma$ multiplier in equation~(\ref{E:Phi}) is significantly lower than unity, making it possible to reach a force equilibrium inside the magnetosphere without surpassing the propeller limit \citep{IS75}, just by decreasing the effective gravity for the optically thin regions. 

\begin{figure}
\adjincludegraphics[width=1.\columnwidth,trim={{0.25\width} 0 0 0},clip]{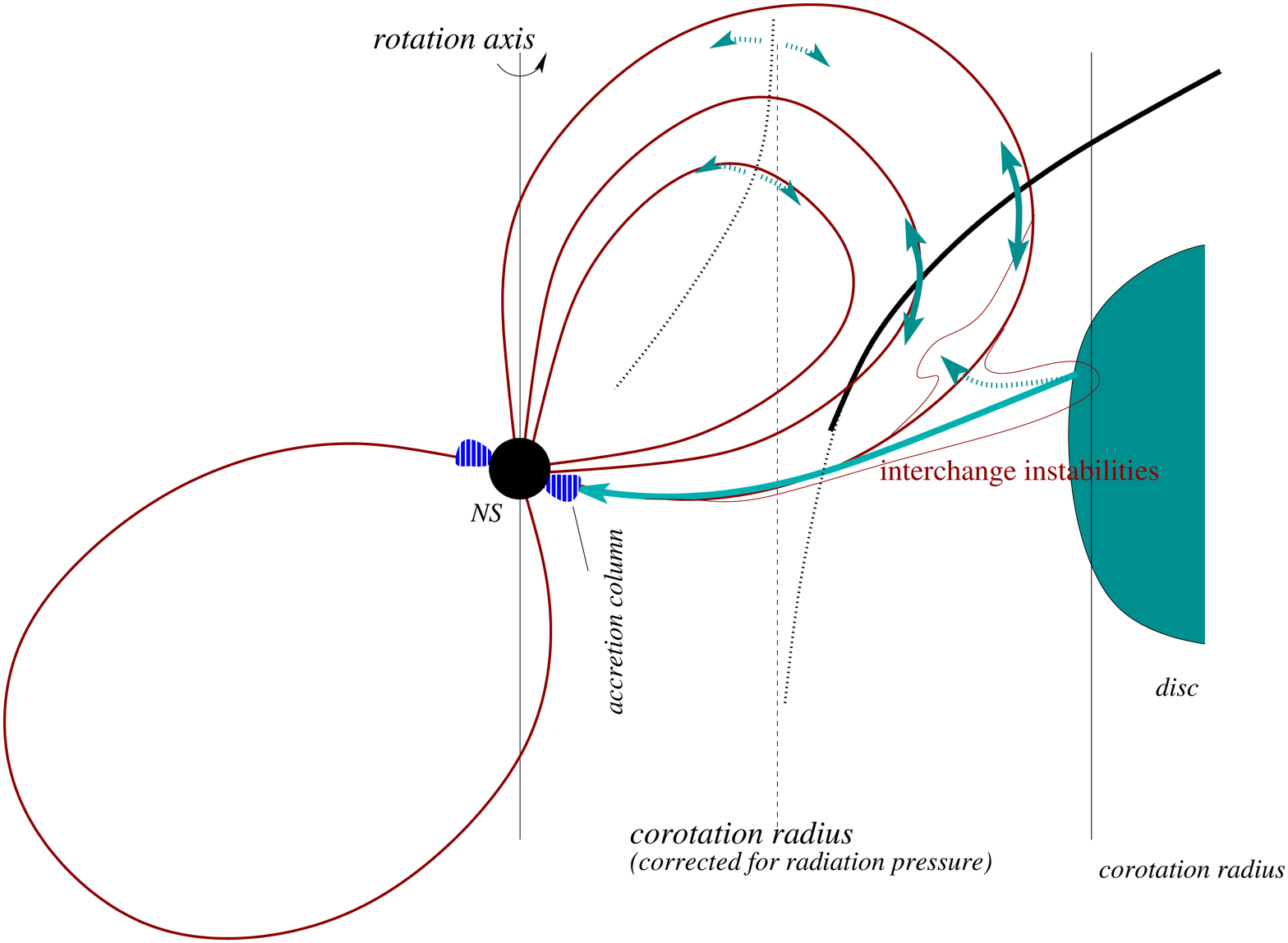}
 \caption{  Qualitative scheme illustrating the structure of the magnetosphere and its equilibrium surfaces. Magnetic field lines are shown with solid red curves. The thick solid black line shows part of the equilibrium surface where the equilibrium points are stable. Each value of $\Gamma$ corresponds to a surface like this, all its spatial dimensions scaling with the Eddington factor as $\propto \left(1-\Gamma \right)^{1/3}$. Blue arrows show the flows of plasma inside the magnetosphere: the main accretion stream (the long single-headed arrow), stable oscillations (solid teal lines), and unstable points (dotted teal).}\label{fig:NSsketch} 
\end{figure}

We suggest that some streams and blobs of accreting matter are at large luminosities trapped in the stable equilibrium points, creating variable reflection and absorption in the observed light curves. An important requirement for the material involved in these motions is its low optical depth to the radiation responsible for the effective anti-gravity. Large optical depth $\tau \gg 1$ will effectively decrease $\Gamma$ by about a factor of $\tau$. 
The matter trapped near the equilibrium points is not necessarily a part of the accretion flow, and thus can have an optical depth much smaller than the optical depth of the accretion flow itself. The probable origin of the oscillating matter is the same as that of the accretion flow: interchange instabilities effectively load some of the closed magnetic field lines with mass. In general, the magnetosphere is transparent to the thermal radiation of the accretion column (otherwise it would be impossible to observe X-ray pulsars), and some portions of plasma will always suffer the action of a $\propto R^{-2}$ radiation force. In Fig.~\ref{fig:NSsketch}, we qualitatively show the flows and equilibrium surfaces of an X-ray pulsar magnetosphere as we see it. 

\subsection{Observability of the inertial oscillation mode}

The easiest way to make the inertial oscillations visible is obscuration of the central source. 
Even at high, super-Eddington accretion rate, accretion column is much more compact than the size of the magnetosphere, that allows any variation of density to block some part of the emission from the column. 
Observed fractional variations of the order percents and tens of percents \citep{revnivtsev09} requires comparable optical depths, $\tau \sim 0.1\div 1$. 
For this, of course, the optical depth of the variable density component should not be too low: expected relative flux variations are of the order of optical depth. 

Our frequency estimates are valid in the force-free regime that implies relatively tenuous medium. 
But it this regime, will it be possible to accumulate a considerable optical depth to make the internal oscillations visible? 
The condition of magnetic stress domination $B^2/4\pi \gg \rho v^2$ implies a restriction on the optical depth
\begin{equation}
    \tau \lesssim \varkappa \rho R \ll \frac{\varkappa B^2 R}{4\pi v^2} \sim \dfrac{\varkappa \mu^2}{4\pi \Omega^2 R^7}.
\end{equation}
As we see, dependence on radius is overwhelmingly strong. If we normalize the radius by the corotation radius $R = x \left( GM/\Omega^2\right)^{1/3}$, and substitute Thomson opacity $\varkappa = \varkappa_{\rm T} \simeq 0.35 {\rm\, cm^{2} \, g^{-1}}$,
\begin{equation}
    \tau \ll \frac{0.5}{x^7} \left(\frac{\mu}{10^{31}\rm G\, cm^3}\right)^{2} \left(\frac{P}{100\rm s}\right)^{4/3}.
\end{equation}
Here, the choice of the spin period and magnetic moment are related to the properties of Be/X-ray systems \citep{raguzova, CP12}. Even for more rapidly spinning and less magnetized pulsars (like \bqcam with a spin period of $4.4$s, showing additional spectral components near the spectral break in outbursts, see \citealt{bqcam}), the strong dependence on radius allows to meet the force-free condition at radii several times smaller than the co-rotation or Alfv\'{e}n radius.

\section{Conclusions}\label{sec:conc}

Motion inside a dipolar magnetosphere has certain equilibrium points, that in some cases  (up to 12 per cent of the total solid angle covered by the equilibrium points) are stable. 
Inertial oscillations in these stable points have frequencies scaling with the spin frequency of the star and roughly homogeneously populating the range from zero to about spin frequency. 

Inertial modes in a rotating magnetosphere can not explain the  higher-frequency noise observed, for instance, in \gro, but can account for the varying shape of the PDS near the break frequency observed during the flares of some X-ray pulsars. 
High luminosity is most likely important as a way to decrease the effective gravity and thus create dynamic equilibrium points inside the magnetosphere. 

\section*{Acknowledgments}

We would like to thank Sergey Tsygankov for interesting discussions about the power-density spectra of X-ray pulsars, and the anonymous referee for valuable comments. 
The 3D plot in this paper was produced with help of {\tt mayavi} software \citep{mayavi} and Sebastian Mueller's {\tt mlabtex} package \url{https://github.com/MuellerSeb/mlabtex}.
The authors acknowledge the support from the Program of development of M.V. Lomonosov Moscow State University (Leading Scientific School `Physics of stars, relativistic objects and galaxies'). AB also thanks for partial support of Russian Government Program of Competitive Growth of Kazan Federal University.

\bibliographystyle{mnras}
\bibliography{mybib}                                              

\label{lastpage}

\end{document}